\documentclass[sigconf,anonymous=False,review=False]{acmart}

\usepackage[shortlabels]{enumitem}
\usepackage{colortbl}
\usepackage{makecell}

\AtBeginDocument{%
  }

\setcopyright{none}

\begin{document}

\title[Stressor Type Matters! --- Exploring Factors Influencing Cross-Dataset Generalizability]{Stressor Type Matters! --- Exploring Factors Influencing Cross-Dataset Generalizability of Physiological Stress Detection}

\author{Pooja Prajod}
\email{pooja.prajod@uni-a.de}
\orcid{0000-0002-3168-3508}
\affiliation{%
  \institution{University of Augsburg}
  \city{Augsburg}
  \country{Germany}
}

\author{Bhargavi Mahesh}
\affiliation{%
  \institution{University of Augsburg}
  \city{Augsburg}
  \country{Germany}
}
\email{bhargavi.mahesh@uni-a.de}

\author{Elisabeth Andr\'e}
\affiliation{%
  \institution{University of Augsburg}
  \city{Augsburg}
  \country{Germany}
}
\email{elisabeth.andre@uni-a.de}

\renewcommand{\shortauthors}{Prajod et al.}

\begin{abstract}

Automatic stress detection using heart rate variability (HRV) features has gained significant traction as it utilizes unobtrusive wearable sensors measuring signals like electrocardiogram (ECG) or blood volume pulse (BVP). However, detecting stress through such physiological signals presents a considerable challenge owing to the variations in recorded signals influenced by factors, such as perceived stress intensity and measurement devices. Consequently, stress detection models developed on one dataset may perform poorly on unseen data collected under different conditions. To address this challenge, this study explores the generalizability of machine learning models trained on HRV features for binary stress detection. Our goal extends beyond evaluating generalization performance; we aim to identify the characteristics of datasets that have the most significant influence on generalizability. We leverage four publicly available stress datasets (WESAD, SWELL-KW, ForDigitStress, VerBIO) that vary in at least one of the characteristics such as stress elicitation techniques, stress intensity, and sensor devices. Employing a cross-dataset evaluation approach, we explore which of these characteristics strongly influence model generalizability. Our findings reveal a crucial factor affecting model generalizability: stressor type. Models achieved good performance across datasets when the type of stressor (e.g., social stress in our case) remains consistent. Factors like stress intensity or brand of the measurement device had minimal impact on cross-dataset performance. Based on our findings, we recommend matching the stressor type when deploying HRV-based stress models in new environments. To the best of our knowledge, this is the first study to systematically investigate factors influencing the cross-dataset applicability of HRV-based stress models.

\end{abstract}




\maketitle

\section{Introduction}

The ability to detect stress in real-time has become increasingly important within the fields of affective computing and human-machine interaction (HMI)~\citep{alberdi2016towards}. Early stress detection offers a valuable tool for promoting well-being and potentially preventing long-term health consequences~\citep{greene2016survey, akmandor2017keep, alberdi2016towards}. Consequently, a growing area of research focuses on developing interactive systems that can not only detect stress but also provide personalized interventions to manage it effectively~\citep{yu2018biofeedback, balcombe2022human}.

Researchers have explored various modalities for stress detection, including psychological tools, behavioral patterns, and physiological signals~\citep{giannakakis2019review, alberdi2016towards}. However, physiological signals are considered more reliable than the other methods as they eliminate certain measurement biases. Moreover, the growing popularity of unobtrusive wearable sensors further facilitates continuous stress monitoring through physiological signals.

Stress manifests differently depending on the context~\citep{alberdi2016towards}. For instance, the stress experienced during an exam likely differs from that of public speaking. While controlled lab settings are often used to develop stress detection models, real-world HMI scenarios are far more diverse. Therefore, assessing the generalizability of these models – their ability to perform well in unseen contexts – is crucial for real-world applications. Cross-dataset evaluation, where a model trained on one dataset is tested on another, is a common approach to assess generalizability. Good cross-dataset performance suggests broader applicability of the model.

The existing literature on stress detection models acknowledges the importance of generalizability, with a few studies exploring cross-dataset evaluations~\citep{vos2023generalizable}. These studies typically investigate the applicability of models across different datasets and sometimes explore combining datasets for improved performance (refer to Section~\ref{sec:lit} for details). While prior works often report limited cross-dataset performance, a key gap exists: a lack of research systematically investigating the factors influencing this limited generalizability.

We address this research gap by conducting extensive cross-dataset evaluations using multiple stress datasets. We train three popular machine learning models - random forest classifier (RFC), support vector machine (SVM), multi-layer perceptron (MLP) - using heart rate variability (HRV) features. We aim to identify the characteristics of these datasets that significantly impact model generalizability. Our findings are crucial for developing stress detection models with broader applicability in HMI systems.

\section{Related Work}
\label{sec:lit}

The publicly available datasets such as WESAD~\citep{schmidt2018introducing} and SWELL-KW~\citep{koldijk2014swell}, led to the development of numerous stress detection models~\citep{can2019stress, haque2024state}. Some of these studies (e.g.,~\citep{bobade2020stress}) focused on comparing multiple models to determine the best model for stress detection. However, many of these comparison studies are trained and evaluated on the same datasets. In this section, we discuss some of the existing works that performed cross-dataset evaluations on stress detection models.

\citeauthor{mishra2020evaluating} \cite{mishra2020evaluating} conducted cross-dataset evaluations across four cognitive stress datasets - two had electrocardiogram (ECG) signals, and the other two had blood volume pulse (BVP) signals. In addition to arithmetic tasks, the ECG datasets had a startle response test and cold pressor task as stressors. The authors trained SVM models using HRV features extracted from the mental stress segments of these datasets. While the ECG-based HRV models performed well in detecting stress in each other's arithmetic tasks, they had a performance drop of 15 - 30\% in predicting stress in the BVP datasets. The authors attributed this drop in performance, despite having the same stressor, to the difference in sensors. They also noticed an approximately 20 - 40\% drop in the performance when detecting overall stress (including startle and cold pressor segments), even within the same datasets. Their findings suggest that the models trained on one type of stressor may not be efficient in detecting other stress responses.

\citeauthor{prajod2022generalizability} \cite{prajod2022generalizability} trained ECG-based deep-learning models and HRV-based shallow models (RFC, SVM, MLP) on WESAD and SWELL-KW datasets. While deep learning models outperformed other methods in within-dataset evaluations, they performed poorly in cross-dataset evaluations (WESAD models tested on SWELL-KW and vice versa). Although the HRV-based models performed better than deep learning models in cross-dataset evaluations, their performances were still low. Moreover, the combining datasets did not improve the model's performance on individual datasets. Similarly, \citeauthor{albaladejo2023evaluating} \cite{albaladejo2023evaluating} trained HRV-based models on the WESAD and SWELL-KW datasets. They also observed poor cross-dataset performances and lack of stress detection improvements by combining datasets.

\citeauthor{benchekroun2023cross} \cite{benchekroun2023cross} trained two HRV-based models (RFC, logistic regression) on the MMSD~\citep{benchekroun2022multi} and UWS~\citep{velmovitsky2021towards} datasets. They tested the MMSD models using the UWS data and found that the f1-scores were 12 – 14\% lower than the UWS models (from within-dataset evaluations). They further noted that the f1-score for stress class was very low (less than 50 \%), meaning the models were not very efficient in detecting stressful instances.

\citeauthor{vos2023ensemble} \cite{vos2023ensemble} trained shallow models (RFC, SVM, XGBoost) using heart rate and electro-dermal activity (EDA) features from the SWELL-KW dataset and evaluated them on WESAD and NEURO~\citep{birjandtalab2016non} datasets. All three models showed poor cross-dataset performances. They also implemented an ensemble model and repeated the cross-dataset evaluation. Although this ensemble model yielded a slight improvement, the performance was still poor (f1-score $<$ 50\%). Furthermore, they trained the ensemble model using a combined dataset (SWELL-KW, NEURO, UBFC-Phys~\citep{sabour2021ubfc}) and evaluated it on WESAD. While the accuracy increased slightly, the f1-score dropped further. Their experiments highlight the challenges of developing a generic stress model.

The above works primarily focused on assessing the generalizability of HRV-based models. However, \citeauthor{liapis2021detection} \cite{liapis2021detection} demonstrated that EDA-based shallow models also struggle with generalizability. They trained their models on the WESAD dataset and then evaluated them on their own dataset. Notably, their dataset contained subtle stress instances, unlike WESAD, implying that the stress intensity might further impact generalizability.

\citeauthor{baird2021evaluation} \cite{baird2021evaluation} compared models trained on speech features from three social stress datasets (FAU-TSST, Ulm-TSST and Reg-TSST). These datasets induced stress following the TSST technique. They predicted cortisol levels as a proxy for stress levels. In cross-dataset evaluations, the trends of predicted cortisol levels were aligned for the models, indicating compatibility between datasets. Due to the dataset compatibility, they suggested that training models using data from both these datasets could result in better-performing models.

Table~\ref{tab:stress_cross_dataset_lit} presents an overview of the existing studies on cross-dataset generalizability of stress detection models. Most studies assess the generalizability of a model to determine if it is applicable in other stress scenarios. Some studies take a step further to evaluate whether combining datasets to train models improves stress detection performances. Although most studies observe low generalizability of stress models, few studies provide insights into plausible factors influencing the models' performance. For example, \citeauthor{mishra2020evaluating} highlighted the poor performance of mental stress models in detecting physical stress. Similarly, \citeauthor{liapis2021detection} noted the difference in stress intensities of the evaluated datasets. \citeauthor{prajod2022generalizability} hinted at multiple factors such as stress intensity and measurement devices that could affect generalizability. However, these studies did not further investigate these factors. Hence, a crucial question remains relatively unexplored - What factors or characteristics of the stress datasets need to match for cross-dataset applicability of models?

\begin{table*}[htpb]
\caption{An overview of the existing works that perform cross-dataset evaluations of their stress models}
\label{tab:stress_cross_dataset_lit}
\renewcommand*{\arraystretch}{1.2}
    \centering \renewcommand\cellalign{c}
    \setcellgapes{2pt}\makegapedcells
        \begin{tabular}{|l | c | c | l |} 
        \hline
         {\large \textbf{Paper}} & {\large \textbf{Input}} & {\large \textbf{Datasets}} & {\large \textbf{Aim}} \\
        \hline
         \citeauthor{mishra2020evaluating} \cite{mishra2020evaluating} & HRV & \makecell{Own datasets (mainly \\mental arithmetic tasks)} & Assess generalizability\\
         \citeauthor{liapis2021detection} \cite{liapis2021detection} & EDA & \makecell{WESAD,\\ Own UX stress dataset} & Assess generalizability\\
         \citeauthor{baird2021evaluation} \cite{baird2021evaluation} & Speech & \makecell{Own TSST datasets} & \makecell[l]{Assess compatibility for combining datasets} \\
         \citeauthor{prajod2022generalizability} \cite{prajod2022generalizability} & Raw ECG, HRV & \makecell{WESAD, SWELL-KW} & \makecell[l]{Assess generalizability, Combine datasets} \\
         \citeauthor{albaladejo2023evaluating} \cite{albaladejo2023evaluating} & HRV & \makecell{WESAD, SWELL-KW} & \makecell[l]{Assess generalizability, Combine datasets} \\
         \citeauthor{benchekroun2023cross} \cite{benchekroun2023cross} & HRV & \makecell{MMSD, UWS} & Assess generalizability \\
         \citeauthor{vos2023ensemble} \cite{vos2023ensemble} & HR, EDA & \makecell{WESAD, SWELL-KW, \\ NEURO, UBFC-Phys} & \makecell[l]{Assess generalizability, Combine datasets}\\
         This work & HRV & \makecell{WESAD, SWELL-KW, \\ ForDigitStress, VerBIO} & \makecell[l]{Assess generalizability, Combine datasets, \\ \textbf{Identify factors influencing generalizability}} \\
        \hline
        \end{tabular}
\end{table*}

\section{Materials and Methods}

\subsection{Datasets}
\label{sec:datasets}

In this study, we focus on binary stress detection (stress vs. no-stress) using HRV features. We leverage four publicly available stress datasets: WESAD~\cite{schmidt2018introducing}, SWELL-KW~\cite{koldijk2014swell}, ForDigitStress~\cite{heimerl2023fordigitstress}, and VerBIO~\cite{yadav2020exploring}. While SWELL-KW contains ECG signals that can be used for extracting HRV, the ForDigitStress dataset has BVP signals for extracting HRV. The WESAD and VerBIO datasets contain both ECG and BVP signals. However, we consider only the BVP signals from the VerBIO dataset to remove some redundant comparisons in our analysis. A brief overview of the four datasets is presented in Table~\ref{tab:stress_datasets}.

\subsubsection{\textbf{WESAD}}
The WESAD dataset is a multimodal stress and affect dataset containing various physiological signals, including ECG, EDA, and BVP. The data from 15 participants were collected using a chest-worn RespiBan and a wrist-worn Empatica E4 device. This investigation utilizes the ECG data recorded by the chest-worn device at 700 Hz.

The participants were subject to three conditions: neutral, amusement, and stress. In the stress condition, the participants experienced social stress induced by the TSST technique. The participants engaged in public speaking and mental arithmetic tasks while being evaluated by a three-member panel. To induce amusement, the participants watched selected funny video clips. The experimental sessions began with the neutral condition, followed by the stress and amusement conditions in alternating order. For each participant, the neutral condition lasted for approximately 20 minutes, the stress condition for 10 minutes, and the amusement condition for around 6.5 minutes.

We focus on stress detection, i.e., distinguishing between stress and no-stress samples. Following the labeling scheme proposed by the dataset creators, data from both neutral and amusement conditions were considered as no-stress samples.

\subsubsection{\textbf{SWELL-KW}}
The SWELL-KW dataset is also a multimodal stress dataset that contains two physiological signals, ECG and EDA. This dataset consists of data from 25 participants who engaged in typical knowledge tasks like writing reports and presentations. The ECG data was collected using the TMSI Mobi device at a sampling rate of 2048 Hz. 

The participants underwent three experimental conditions: neutral, email interruptions, and time pressure. During the email interruption session, participants received eight emails, many irrelevant and some requiring responses. In the time pressure condition, participants had to complete the tasks within two-thirds of the allotted neutral session time. Like the WESAD dataset, the first session was always neutral, followed by the other two conditions in alternating order. The neutral and email interruption sessions lasted approximately 45 minutes, while the time pressure session was around 30 minutes long.

Notably, the participants did not report experiencing high stress in any of the three conditions. However, they indicated a higher temporal demand during the time pressure session. While training stress detection models, the dataset creators considered the data from email interruptions and time pressure sessions as stress samples and the neutral session as no-stress samples~\citep{koldijk2016detecting}. Therefore, we follow the same labeling scheme for consistency. However, three participants were excluded due to missing data.

\subsubsection{\textbf{ForDigitStress}}

The ForDigitStress dataset represents another multimodal stress dataset with various behavioral (facial expression, body pose, etc.) and physiological (BVP, EDA) signals. The dataset was collected from 40 participants who attended a mock job interview session. The BVP data was collected using IOM-biofeedback device, operating at 27 Hz.

The experimental session was divided into three phases: preparation, interview, and post-interview. The participants were asked to submit their resumes in advance so that the interviewer could customize the questions depending on the participant. During the interview phase, the interviewer questioned the participants on topics such as their strengths/weaknesses, salary expectations, and hypothetical job-related scenarios. The interview phase lasted for about 25 mins. In addition to self-reported stress levels, the participants' saliva samples were collected for assessing the cortisol levels. The cortisol levels served as ground truths for the presence of stress.

Although both preparation and post-interview phases can be considered as no-stress conditions, the authors suggest using data from later parts of the post-interview phase based on the cortisol levels. So, we utilize the interview phase as stress samples and the last segments of post-interview phase (15 - 20 mins) as no-stress samples.

\subsubsection{\textbf{VerBIO}}
The VerBIO dataset was collected to investigate if exposing participants to public speaking through virtual reality (VR) training would reduce the public speaking anxiety.The dataset contains data collected from 55 participants who were recruited for two real and eight virtual oral presentations over two days. The physiological data from two wearable sensors, Empatica E4 (BVP, EDA, skin temperature) and Actiwave Cardio Monitor (ECG) was acquired. Several self-assessments questionnaires were employed to capture demographics, state- and trait-based psychological measures.

The experiment was divided into three sessions: pre-interview (Pre), interview (Test), and post-interview (Post) sessions. The Pre and Post sessions involved presenting in front of real audience, whereas the Test session involved VR audience. Each session further constituted relaxation, preparation, and presentation phases. The presentation topics were of general interest and were about four minutes long.

We note that the participation dropped through the three sessions, with maximum participants in the Pre phase and lowest in Post phase. To avoid drastic imbalances in data between participants, we utilized only the data from the Pre session. We labeled the data belonging to relaxation phases as no-stress and those of presentation phase as stressful. In our analysis, we normalize the data for each participant using few minutes of no-stress data (see Section~\ref{sec:data_processing}). Hence, 10 participants with low amount of no-stress data were excluded.

\begin{table*}[ht]
    \renewcommand*{\arraystretch}{1.2}
    \centering \renewcommand\cellalign{l}
    \setcellgapes{2pt}\makegapedcells
    \caption{An overview of some key characteristics of the four stress datasets}
    \label{tab:stress_datasets}
        \begin{tabular}{|p{6.5em} |p{10.5em} |p{10.5em} | p{10.5em} | p{10.5em} |} 
        \hline
          & \textbf{WESAD} & \textbf{SWELL-KW} & \textbf{ForDigitStress} & \textbf{VerBIO} \\
        \hline
         Stressor & TSST & \makecell[l]{Interruptions,\\ Time pressure} & Job interview & Public speaking \\
         Stressor type & Social & Cognitive & Social & Social \\
         Sensor & \makecell[l]{ECG: RespiBan, 700 Hz,\\ BVP: Empatica E4, 64 Hz} & ECG: TMSI Mobi, 2048 Hz & BVP: IOM, 27 Hz & BVP: Empatica E4, 64 Hz \\
         Avg. stress level & 18.5/24 (STAI) & 3.5/10 (Likert scale) & \makecell[l]{5.4/10 (Likert scale),\\ 6.5/10 (Cortisol)} & \\
         Participants & 15 & 22 & 40 & 45 \\
         \makecell[l]{Data duration\\ (per participant)} & \makecell[l]{stress: 10 mins,\\ no-stress: 26.5 mins} & \makecell[l]{stress: 75 mins,\\ no-stress: 45 mins} & \makecell[l]{stress: 25 mins,\\ no-stress: 15 - 20 mins} & \makecell[l]{stress: 2 - 5 mins,\\ no-stress: 4 - 6 mins}\\
        \hline
        \end{tabular}
\end{table*}

\subsection{Approach}
Our investigations involve the three assessments listed below. The idea is to run a series of these assessments using the datasets described in Section~\ref{sec:datasets} to investigate the dataset characteristics that considerably influence model generalizability.

\begin{enumerate}
    \item \textbf{Within-dataset Assessment:} involves training and evaluating models on the same dataset. We utilize the leave-one-subject-out (LOSO) technique to evaluate the models' performance on unseen participants of the same dataset.

    \item \textbf{Cross-dataset Assessment:} involves evaluating the models trained on one dataset using another dataset. This evaluation assesses to what extent these models can detect stress in new participants in different settings.

    \item \textbf{Combining Datasets:} involves training new models on a combined dataset consisting of data from two or more stress datasets, again using the LOSO technique. This step investigates potential improvements in models’ performances due to an increase in the number and variations of the training data.
    
\end{enumerate}

While ECG-based and BVP-based HRVs reflect similar physiological information related to heart activity, the models trained on these features might not perform equally well~\cite {gupta2023multimodal}. This discrepancy can be attributed to the BVP signals being more prone to noise from body movements than ECG~\cite {martinho2018towards}. This noise can affect the signal quality, which can, in turn, impact the stress detection performance. To avoid such influences in cross-dataset assessment, the above assessments are followed for ECG-based and BVP-based HRV models separately.

\subsection{Data Processing}
\label{sec:data_processing}
\subsubsection{\textbf{ECG signals}}

\begin{figure}[ht]
  \centering
  \includegraphics[width=0.8\linewidth]{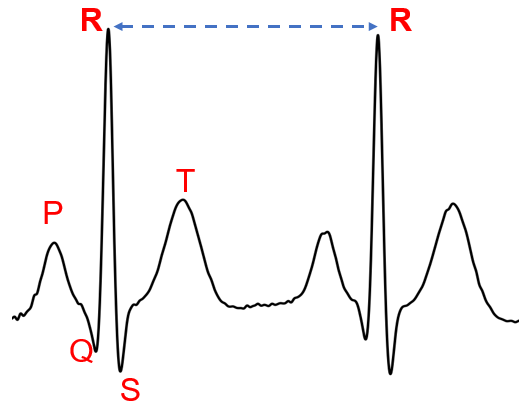}
  \caption{An example plot of ECG signal, marked with various repeated components of the signal.}
  \label{fig:ecg_marked}
  \Description{A woman and a girl in white dresses sit in an open car.}
\end{figure}

The ECG signals are typically sampled at high frequencies and contain noises such as baseline wander (low-frequency, 0.5 - 0.6 Hz) and powerline interference (50 or 60 Hz). Figure~\ref{fig:ecg_marked} illustrates two beats from an ECG signal. Computing HRV relies on detecting the R peaks in the QRS complex of the ECG signal. We applied a second-order Butterworth band-pass filter with a frequency band of 8 - 20 Hz for optimal QRS signal-to-noise ratio \cite{elgendi2010frequency}.

For R-peak detection, we utilized the algorithm proposed by \citep{elgendi2010frequency}. This algorithm is based on two key assumptions for healthy adults:

\begin{enumerate}[(a)]
    \item A QRS complex contains one and only one heartbeat
    \item The duration of a typical QRS complex is in the range of 80 - 120 milliseconds
\end{enumerate}

\subsubsection{\textbf{BVP signals}}

\begin{figure}[ht]
  \centering
  \includegraphics[width=0.6\linewidth]{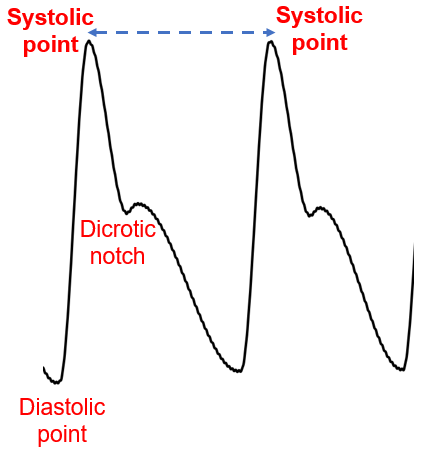}
  \caption{An example plot of BVP signal, marked with various repeated components of the signal.}
  \label{fig:bvp_marked}
  \Description{A woman and a girl in white dresses sit in an open car.}
\end{figure}

Like ECG, the BVP signal is susceptible to baseline wander and high-frequency noise. Hence, a band-pass filter (0.5 - 8 Hz) was applied to reduce these noises~\citep{elgendi2013systolic}. Figure~\ref{fig:bvp_marked} illustrates two beats from a BVP signal. To derive the HRV signal from the BVP, the systolic peaks had to be detected. For this purpose, we employed the peak-finding algorithm from \cite{heimerl2023fordigitstress} to detect points that meet the following criteria:

\begin{enumerate}[(a)]
    \item Amplitude threshold: The peaks had to be taller than a certain threshold. This threshold was set based on the distribution of peak heights in the entire signal.
    \item Distance between peaks: To avoid identifying every fluctuation as a peak, consecutive peaks had to be separated by a minimum interval of 0.333 seconds. This value corresponds to a maximum heart rate of 3 beats per second (180 beats per minute).
\end{enumerate}

\subsubsection{\textbf{HRV Features}}
Once the R-peaks (for ECG) or systolic peaks (for BVP) were identified, the time intervals between successive peaks were calculated to form the HRV signals. The features were calculated using 60-second segments with 59 seconds of overlap between consecutive segments.

A total of 22 well-known features~\citep{schmidt2018introducing, pham2021heart, heimerl2023fordigitstress, prajod11flow} were computed from the extracted HRV signals. These features belonged to the time domain (13 features), frequency domain (5 features), and poincaré plot characteristics (4 features). 
These features were calculated using the NeuroKit2 Python library~\citep{makowski2021neurokit2}.

The sensors used in the different datasets differ, potentially resulting in values recorded on different scales. In addition, the range of physiological recordings may vary from participant to participant~\cite{braithwaite2013guide, nkurikiyeyezu2019effect, sarkar2020self}. To mitigate the effect of these differences, participant-specific Min-Max normalization was applied to each HRV feature. For real-time stress detection, the entire dataset would not be available for normalization. Similar to~\cite{luong2020towards, prajod2022generalizability}, we used 5 minutes of neutral data to compute normalization parameters (minimum and maximum values) for each participant. For the VerBIO dataset, we used neutral data for normalization because many of the participants had less than 5 minutes of neutral data.

\subsection{Machine Learning Models}
We trained the following three machine learning models using the extracted HRV features, following the LOSO procedure. To account for the imbalanced sample distribution of the datasets, the ``class\_weight" hyperparameter for all models was set inversely proportional to the sample frequencies.

\subsubsection{\textbf{Random Forest Classifier (RFC)}}

This is an ensemble learning method that combines predictions from multiple decision trees for improved performance and reduced overfitting. Each tree is trained on a subset of the available training set. The final prediction is determined by aggregating the predictions from all the trees (e.g., majority vote). This strategy often results in a better performance, even if the individual decision trees are weak predictors. A total of 200 decision trees (also called estimators) were trained, with a maximum depth of 5 for each tree. 

\subsubsection{\textbf{Support Vector Machine (SVM)}}

This is a commonly used supervised learning method for binary classification tasks. During the training process of this model, the objective is to find a hyperplane within the feature space that separates the data points belonging to different classes. We utilized a linear SVM classifier.

\subsubsection{\textbf{Multi-layer Perceptron (MLP)}}
This is a simple feed-forward neural network (also called simple artificial neural network), which has been growing in popularity for stress detection~\citep{bobade2020stress, zawad2023hybrid, albaladejo2023evaluating}. Our implementation followed an architecture consisting of an input layer, two hidden layers, and a prediction layer. The input layer received data represented as the normalized HRV features. A dropout layer (rate $=$ 0.2) was included after the input layer to mitigate overfitting of the model. The two hidden layers (ReLU activation) followed the dropout layer, with 12 nodes for the first hidden layer and 6 nodes for the second hidden layer. This final layer outputs the classification result using a Sigmoid activation function.

This model was trained using the SGD optimizer (learning rate $=$ 0.001) and weighted loss. It was also trained in batches of 256 samples. We utilized the early stopping technique, where the training stopped if the validation loss did not decrease for 15 consecutive iterations.

\section{Results}

We employed three evaluation strategies to assess model performances: within-dataset, cross-dataset, and combined dataset evaluations. For all evaluations, accuracy and f1-score metrics were used to quantify model performance. The detailed results for each evaluation strategy are presented below.

\subsection{Within-dataset Assessment}

We employed LOSO technique to train and evaluate model performance within each of the four stress datasets. The average f1-score and accuracy for each dataset are presented in Table~\ref{tab:within_loso}.

\begin{table}[thpb]
    \renewcommand*{\arraystretch}{1.2}
    \centering
    \caption{Results of within-dataset LOSO evaluation of HRV models conducted on the four stress datasets}
    \label{tab:within_loso}
    \begin{tabular}{|c | c | c|}
        \hline 
        {\large \textbf{Model}} & {\large \textbf{F1-score}} & {\large \textbf{Accuracy}}\\
        \hline
        \multicolumn{3}{|c|}{\cellcolor[rgb]{0.863, 0.863, 0.863} Test on SWELL-KW, ECG} \\
        \hline
        SWELL-KW baseline~\citep{koldijk2016detecting} & - & 0.589 \\
        SWELL-KW RFC & 0.634 & 0.664 \\
        SWELL-KW SVM & 0.597 & 0.626 \\
        SWELL-KW MLP & \textbf{0.667} & \textbf{0.692} \\
        \hline
        \multicolumn{3}{|c|}{\cellcolor[rgb]{0.863, 0.863, 0.863} Test on WESAD, ECG} \\
        \hline
        WESAD (ECG-HRV) baseline~\citep{schmidt2018introducing} & 0.813 & 0.854 \\
        WESAD (ECG-HRV) RFC & 0.819 & 0.863 \\
        WESAD (ECG-HRV) SVM & 0.814 & 0.862 \\
        WESAD (ECG-HRV) MLP & \textbf{0.844} & \textbf{0.880} \\
        \hline
        \multicolumn{3}{|c|}{\cellcolor[rgb]{0.863, 0.863, 0.863} Test on WESAD, BVP} \\
        \hline
        WESAD (BVP-HRV) baseline~\citep{schmidt2018introducing} & 0.830 & 0.858 \\
        WESAD (BVP-HRV) RFC & 0.768 & 0.826 \\
        WESAD (BVP-HRV) SVM & 0.763 & 0.792 \\
        WESAD (BVP-HRV) MLP & \textbf{0.780} & \textbf{0.829} \\
        \hline
        \multicolumn{3}{|c|}{\cellcolor[rgb]{0.863, 0.863, 0.863} Test on ForDigitStress, BVP} \\
        \hline
        ForDigitStress baseline~\citep{heimerl2023fordigitstress} & 0.784 & 0.797 \\
        ForDigitStress RFC & 0.810 & 0.829 \\
        ForDigitStress SVM & 0.787 & 0.811 \\
        ForDigitStress MLP & \textbf{0.814} & \textbf{0.831} \\
        \hline
        \multicolumn{3}{|c|}{\cellcolor[rgb]{0.863, 0.863, 0.863} Test on VerBIO, BVP} \\
        \hline
        VerBIO baseline~\citep{yadav2020exploring} & - & - \\
        VerBIO RFC & \textbf{0.949} & \textbf{0.962} \\
        VerBIO SVM & 0.879 & 0.904 \\
        VerBIO MLP & 0.924 & 0.939 \\
        \hline
    \end{tabular}
\end{table}

The VerBIO dataset yielded the best overall performance, with models achieving average accuracies and f1-scores exceeding 90\%. On the other hand, models trained on the SWELL-KW dataset exhibited lower average performance, with f1-scores and accuracies ranging from 60\% to 70\%. All other datasets achieved average f1-scores above 75\% and average accuracies greater than 80\%.

Except for VerBIO, MLP consistently achieved the highest performance within each dataset, followed by RFC and then SVM. For VerBIO, RFC outperformed all other models. Notably, all our models surpassed the baselines established in the respective dataset papers, except for the WESAD BVP-HRV models.  The performance differences between models were relatively small, typically within a 5\% margin.

Within the WESAD dataset, models trained on ECG-derived HRV features outperformed those using BVP-HRV features. However, the WESAD dataset paper~\cite{schmidt2018introducing} reported slightly higher performance for BVP-HRV models.

\subsection{Cross-dataset Assessment}

To assess the generalizability of the models, we conducted cross-dataset evaluations. Each LOSO model trained on one dataset was tested on unseen data from another dataset. For ECG-derived HRV features, we tested the SWELL-KW models on data from the WESAD dataset, and vice versa. The results for these cross-dataset evaluations are presented in Table~\ref{tab:swell_cross} (SWELL-KW) and Table~\ref{tab:wesad_ecg_cross} (WESAD, ECG).

\begin{table}[htbp]
    \caption{Cross-dataset evaluation of SWELL-KW models on WESAD (ECG-HRV) dataset}
    \label{tab:swell_cross}
    \renewcommand*{\arraystretch}{1.2}
    \centering
    \begin{tabular}{|c|c|c|}
        \hline
        {\large \textbf{Model}} & {\large \textbf{F1-score}} & {\large \textbf{Accuracy}}\\
        \hline
        \multicolumn{3}{|c|}{\cellcolor[rgb]{0.863, 0.863, 0.863} Test on WESAD (ECG-HRV)} \\
        \hline
        SWELL-KW RFC & \textbf{0.557} & \textbf{0.676} \\
        SWELL-KW SVM & 0.357 & 0.482 \\
        SWELL-KW MLP & 0.427 & 0.574 \\
        \hline
    \end{tabular}
\end{table}

The SWELL-KW models performed poorly in the cross-dataset evaluation using WESAD ECG-HRV data. The RFC models achieved the best average performance with an f1-score of 55.7\% and an accuracy of 67.6\%. These values are considerably lower than the worst-performing within-dataset WESAD (ECG-HRV) model, which achieved f1-score $=$ 81.4\% and accuracy $=$ 86.2\%.

\begin{table}[htbp]
    \renewcommand*{\arraystretch}{1.2}
    \centering
    \caption{Cross-dataset evaluation of WESAD (ECG-HRV) models on SWELL-KW dataset}
    \label{tab:wesad_ecg_cross}
    \begin{tabular}{|c|c|c|}
        \hline
        {\large \textbf{Model}} & {\large \textbf{F1-score}} & {\large \textbf{Accuracy}}\\
        \hline
        \multicolumn{3}{|c|}{\cellcolor[rgb]{0.863, 0.863, 0.863} Test on SWELL-KW} \\
        \hline
        WESAD (ECG-HRV) RFC & 0.432 & 0.450 \\
        WESAD (ECG-HRV) SVM & 0.421 & 0.439 \\
        WESAD (ECG-HRV) MLP & \textbf{0.468} & \textbf{0.479} \\
        \hline
    \end{tabular}
\end{table}

Like SWELL-KW models, the WESAD ECG-HRV models also performed poorly in cross-dataset evaluation. While MLP achieved the highest average performance in this cross-dataset evaluation, all models yielded f1-scores and accuracies below 50\%. 

We employed a similar cross-dataset evaluation approach for models trained on BVP-derived HRV features. Each LOSO model was tested on unseen data from two different datasets. For instance, the WESAD (BVP-HRV) models were evaluated using data from the ForDigitStress and VerBIO datasets. The results of cross dataset evaluation for WESAD (BVP-HRV), ForDigitStress, and VerBIO models are presented in Tables~\ref{tab:wesad_bvp_cross}, \ref{tab:fordigitstress_cross}, and \ref{tab:verbio_cross}, respectively.

\begin{table}[htbp]
    \renewcommand*{\arraystretch}{1.2}
    \centering
    \caption{Cross-dataset evaluation of WESAD (BVP-HRV) models on ForDigitStress and VerBIO datasets}
    \label{tab:wesad_bvp_cross}
    \begin{tabular}{|c|c|c|}
        \hline
        {\large \textbf{Model}} & {\large \textbf{F1-score}} & {\large \textbf{Accuracy}}\\
        \hline
        \multicolumn{3}{|c|}{\cellcolor[rgb]{0.863, 0.863, 0.863} Test on ForDigitStress} \\
        \hline
        WESAD (BVP-HRV) RFC & 0.731 & 0.740 \\
        WESAD (BVP-HRV) SVM & \textbf{0.774} & \textbf{0.775} \\
        WESAD (BVP-HRV) MLP & 0.740 & 0.744 \\
        \hline
        \multicolumn{3}{|c|}{\cellcolor[rgb]{0.863, 0.863, 0.863} Test on VerBIO} \\
        \hline
        WESAD (BVP-HRV) RFC & \textbf{0.904} & \textbf{0.907} \\
        WESAD (BVP-HRV) SVM & 0.873 & 0.877 \\
        WESAD (BVP-HRV) MLP & 0.888 & 0.893 \\
        \hline
    \end{tabular}
\end{table}

The models trained on WESAD BVP-HRV features showed good cross-dataset performance on ForDigitStress and VerBIO datasets. SVM achieved the best average f1-score and accuracy on the ForDigitStress dataset, whereas RFC was better in the VerBIO dataset. The performance drop compared to the best within-dataset models on ForDigitStress and VerBIO was minimal, ranging from 4\% to 6\%.

\begin{table}[htbp]
    \renewcommand*{\arraystretch}{1.2}
    \centering
    \caption{Cross-dataset evaluation of ForDigitStress models on WESAD (BVP-HRV) and VerBIO datasets}
    \label{tab:fordigitstress_cross}
    \begin{tabular}{|c|c|c|}
        \hline
        {\large \textbf{Model}} & {\large \textbf{F1-score}} & {\large \textbf{Accuracy}}\\
        \hline
        \multicolumn{3}{|c|}{\cellcolor[rgb]{0.863, 0.863, 0.863} Test on WESAD (BVP-HRV)} \\
        \hline
        ForDigitStress RFC & \textbf{0.789} & \textbf{0.820} \\
        ForDigitStress SVM & 0.779 & 0.812 \\
        ForDigitStress MLP & 0.763 & 0.810 \\
        \hline
        \multicolumn{3}{|c|}{\cellcolor[rgb]{0.863, 0.863, 0.863} Test on VerBIO} \\
        \hline
        ForDigitStress RFC & \textbf{0.856} & \textbf{0.865} \\
        ForDigitStress SVM & 0.836 & 0.846 \\
        ForDigitStress MLP & 0.784 & 0.805 \\
        \hline
    \end{tabular}
\end{table}

The ForDigitStress models performed well on data from WESAD (BVP-HRV) and VerBIO datasets, with all models achieving f1-scores higher than 75\% and more than 80\% accuracy. The RFC showed the best performance on both external datasets.

\begin{table}[htbp]
    \renewcommand*{\arraystretch}{1.2}
    \centering
    \caption{Cross-dataset evaluation of VerBIO models on WESAD (BVP-HRV) and ForDigitStress datasets}
    \label{tab:verbio_cross}
    \begin{tabular}{|c|c|c|}
        \hline
        {\large \textbf{Model}} & {\large \textbf{F1-score}} & {\large \textbf{Accuracy}}\\
        \hline
        \multicolumn{3}{|c|}{\cellcolor[rgb]{0.863, 0.863, 0.863} Test on WESAD (BVP-HRV)} \\
        \hline
        VerBIO RFC & 0.744 & 0.759 \\
        VerBIO SVM & 0.769 & 0.789 \\
        VerBIO MLP & \textbf{0.780} & \textbf{0.803} \\
        \hline
        \multicolumn{3}{|c|}{\cellcolor[rgb]{0.863, 0.863, 0.863} Test on ForDigitStress} \\
        \hline
        VerBIO RFC & \textbf{0.823} & \textbf{0.823} \\
        VerBIO SVM & 0.684 & 0.688 \\
        VerBIO MLP & 0.747 & 0.752 \\
        \hline
    \end{tabular}
\end{table}

The VerBIO models exhibited good generalizability to unseen data from both WESAD (BVP-HRV) and ForDigitStress datasets. A notable exception was the SVM models tested on ForDigitStress achieved f1-scores slightly below 70\%, representing a drop of 10 - 12\% compared to the lowest performing within-dataset ForDigitStress model. Interestingly, the VerBIO MLP - which performed the best on ForDigitStress data - achieved a slightly better performance than the within-dataset ForDigitStress models.

\subsection{Combining Datasets}
To explore the potential benefits of combining datasets, we trained additional models using the LOSO technique. The SWELL-KW and WESAD (ECG-HRV) data were combined to train ECG-derived HRV models. Similarly, a combined dataset consisting of WESAD (BVP-HRV), ForDigitStress, and VerBIO data was used to train BVP-derived HRV models. The LOSO results for combined ECG-based and BVP-based HRV models are presented in Tables~\ref{tab:ecg_comb} and \ref{tab:bvp_comb}, respectively.

\begin{table}[htbp]
    \renewcommand*{\arraystretch}{1.2}
    \centering
    \caption{Results of LOSO evaluation of ECG-derived HRV models trained by combining data from SWELL-KW and WESAD (ECG-HRV) datasets}
    \label{tab:ecg_comb}
    \begin{tabular}{|c|c|c|}
        \hline
        {\large \textbf{Model}} & {\large \textbf{F1-score}} & {\large \textbf{Accuracy}}\\
        \hline
        \multicolumn{3}{|c|}{\cellcolor[rgb]{0.863, 0.863, 0.863} Test on SWELL-KW} \\
        \hline
        Combined ECG-HRV RFC & 0.644 & 0.671 \\
        Combined ECG-HRV SVM & 0.587 & 0.615 \\
        Combined ECG-HRV MLP & \textbf{0.660} & 0.\textbf{679} \\
        \hline
        \multicolumn{3}{|c|}{\cellcolor[rgb]{0.863, 0.863, 0.863} Test on WESAD (ECG-HRV)} \\
        \hline
        Combined ECG-HRV RFC & 0.679 & 0.761 \\
        Combined ECG-HRV SVM & 0.526 & 0.652 \\
        Combined ECG-HRV MLP & \textbf{0.718} & \textbf{0.792} \\
        \hline
        \multicolumn{3}{|c|}{\cellcolor[rgb]{0.863, 0.863, 0.863} Combined ECG-HRV Results} \\
        \hline
        Combined ECG-HRV RFC & 0.658 & 0.707 \\
        Combined ECG-HRV SVM & 0.587 & 0.615 \\
        Combined ECG-HRV MLP & \textbf{0.683} & \textbf{0.725} \\
        \hline
    \end{tabular}
\end{table}

Among the models trained on combined SWELL-KW and WESAD (ECG-HRV), MLP outperformed others on both datasets. However, the performances on each dataset were lower than the within-dataset models. While the drop in performance for SWELL-KW dataset was relatively small, there was around 12\% drop in f1-score and 9\% drop in accuracy for WESAD dataset. 

\begin{table}[htbp]
    \renewcommand*{\arraystretch}{1.2}
    \centering
    \caption{Results of LOSO evaluation of BVP-derived HRV models trained by combining data from WESAD (BVP-HRV), ForDigitStress, and VerBIO datasets}
    \label{tab:bvp_comb}
    \begin{tabular}{|c|c|c|}
        \hline
        {\large \textbf{Model}} & {\large \textbf{F1-score}} & {\large \textbf{Accuracy}}\\
        \hline
        \multicolumn{3}{|c|}{\cellcolor[rgb]{0.863, 0.863, 0.863} Test on WESAD (BVP-HRV)} \\
        \hline
        Combined BVP-HRV RFC & 0.785 & 0.816 \\
        Combined BVP-HRV SVM & 0.768 & 0.813 \\
        Combined BVP-HRV MLP & \textbf{0.823} & \textbf{0.863} \\
        \hline
        \multicolumn{3}{|c|}{\cellcolor[rgb]{0.863, 0.863, 0.863} Test on ForDigitStress} \\
        \hline
        Combined BVP-HRV RFC & 0.809 & 0.828 \\
        Combined BVP-HRV SVM & 0.776 & 0.800 \\
        Combined BVP-HRV MLP & \textbf{0.811} & \textbf{0.831} \\
        \hline
        \multicolumn{3}{|c|}{\cellcolor[rgb]{0.863, 0.863, 0.863} Test on VerBIO} \\
        \hline
        Combined BVP-HRV RFC & \textbf{0.909} & \textbf{0.934} \\
        Combined BVP-HRV SVM & 0.845 & 0.886 \\
        Combined BVP-HRV MLP & 0.880 & 0.913 \\
        \hline
        \multicolumn{3}{|c|}{\cellcolor[rgb]{0.863, 0.863, 0.863} Combined BVP-HRV Results} \\
        \hline
        Combined BVP-HRV RFC & \textbf{0.850} & 0.\textbf{874} \\
        Combined BVP-HRV SVM & 0.806 & 0.840 \\
        Combined BVP-HRV MLP & 0.844 & 0.873 \\
        \hline
    \end{tabular}
\end{table}

Combining WESAD (BVP-HRV), ForDigitStress, and VerBIO datasets resulted in models with good stress detection performance across all three datasets. The best average f1-scores and accuracies for each dataset was greater than 80\%. MLP trained on the combined dataset outperformed the best within-dataset WESAD (BVP-HRV) model by around 4\%. The best performance of combined models on the ForDigitStress dataset was similar to the within-dataset results. However, in the VerBIO dataset, the highest average performance was 3 - 4\% lower than the best within-dataset performance.

\section{Discussion}

The MLP models achieved the best results in most within-dataset evaluations. Our observation aligns with the findings of \cite{bobade2020stress, prajod2022generalizability, albaladejo2023evaluating}, where a simple feed-forward network achieved better performance than other machine learning methods such as SVM and RFC. We also observed that RFC performed better in many cross-dataset evaluations. However, this trend is not very consistent across datasets.

Cross-dataset evaluations revealed significant limitations in generalizability for models trained on SWELL-KW and WESAD (ECG-HRV) datasets. Combining these datasets did not improve performance and, in the case of WESAD models, even led to a decline. This observation highlights the importance of data compatibility when considering such strategies. Our findings regarding the cross-dataset performance and combining datasets with respect to SWELL-KW and WESAD datasets are in line with the observations of previous works~\citep{prajod2022generalizability, albaladejo2023evaluating}. 

As highlighted in Table~\ref{tab:stress_datasets}, the SWELL-KW and WESAD (ECG-HRV) datasets differ in many factors including stressors, experienced stress intensity, and the measurement devices. To pin-point the dataset characteristics which considerably influences the cross-dataset performance, we considered two additional datasets: ForDigitStress and VerBIO. Cross-dataset evaluations were conducted using these additional datasets and WESAD. The WESAD and VerBIO datasets share more similarities than the ForDigitStress dataset. Both datasets utilized the Empatica E4 device to measure the BVP signals. While WESAD employed TSST to elicit stress, VerBIO relied on public speaking task that is a sub-task of the TSST protocol. On the other hand, the ForDigitStress dataset utilized a different measurement device and employed a mock job interview to induce stress. However, we note that the three datasets rely on social evaluation as a primary source of stress.

The models trained on these three datasets (WESAD, VerBIO, ForDigitStress) exhibited good cross-dataset performance. This suggests that factors like brand of measurement device, stress-elicitation technique, and even stress intensity (within a reasonable range) may not significantly impact generalizability when the stressor type remains consistent (social stress in this case). Together with the findings of \citep{mishra2020evaluating} and \citep{baird2021evaluation} - where good cross-dataset performances were observed in datasets involving virtually same tasks (mental arithmetic and TSST, respectively) - we infer that stressor type plays a crucial role in cross-dataset generalizability of stress models.

While our results suggest stress intensity may not be a critical factor within a reasonable range, it warrants further exploration. The low stress intensity in the SWELL-KW dataset might have contributed to the poor performance of WESAD models trained on high-intensity stress data. Interestingly, SWELL-KW models, designed for low-intensity stress detection, also struggled with high-intensity stress from WESAD data.

Overall, this study highlights the importance of considering stressor type and data compatibility when developing generalizable stress detection models.

\section{Conclusion}

Stress detection models with broader applicability are crucial due to the diverse nature of stress experiences across various scenarios. Identifying factors that influence cross-dataset generalizability is essential for achieving this goal. This study addressed this gap by conducting cross-dataset evaluations on four datasets (SWELL-KW, WESAD, ForDigitStress, VerBIO), which contains both shared and distinct characteristics. We trained HRV-based machine learning models (RFC, SVM, MLP) using ECG or BVP signals from these datasets. Our key finding is that stressor type is the most prominent factor influencing cross-dataset applicability. Models trained on datasets with similar stressor types exhibited good generalizability, while those with different stressors showed lower performance. Conversely, factors like stress elicitation method and stress intensity had a minimal impact within the explored datasets. Furthermore, matching stressor type proved crucial for enhancing stress detection performance when combining datasets.

This study focused on ECG- and BVP-derived HRV features. In future works, we will explore whether these findings extend to other stress-related modalities (e.g., EDA). Additionally, a more extensive evaluation involving a wider range of datasets encompassing different stress types (e.g., physical stress) would further validate these observations.


\bibliographystyle{ACM-Reference-Format}
\bibliography{references}


\begin{thebibliography}{38}


\ifx \showCODEN    \undefined \def \showCODEN     #1{\unskip}     \fi
\ifx \showDOI      \undefined \def \showDOI       #1{#1}\fi
\ifx \showISBNx    \undefined \def \showISBNx     #1{\unskip}     \fi
\ifx \showISBNxiii \undefined \def \showISBNxiii  #1{\unskip}     \fi
\ifx \showISSN     \undefined \def \showISSN      #1{\unskip}     \fi
\ifx \showLCCN     \undefined \def \showLCCN      #1{\unskip}     \fi
\ifx \shownote     \undefined \def \shownote      #1{#1}          \fi
\ifx \showarticletitle \undefined \def \showarticletitle #1{#1}   \fi
\ifx \showURL      \undefined \def \showURL       {\relax}        \fi
\providecommand\bibfield[2]{#2}
\providecommand\bibinfo[2]{#2}
\providecommand\natexlab[1]{#1}
\providecommand\showeprint[2][]{arXiv:#2}

\bibitem[Akmandor and Jha(2017)]%
        {akmandor2017keep}
\bibfield{author}{\bibinfo{person}{Ayten~Ozge Akmandor} {and} \bibinfo{person}{Niraj~K Jha}.} \bibinfo{year}{2017}\natexlab{}.
\newblock \showarticletitle{Keep the stress away with SoDA: Stress detection and alleviation system}.
\newblock \bibinfo{journal}{\emph{IEEE Transactions on Multi-Scale Computing Systems}} \bibinfo{volume}{3}, \bibinfo{number}{4} (\bibinfo{year}{2017}), \bibinfo{pages}{269--282}.
\newblock


\bibitem[Albaladejo-Gonz{\'a}lez et~al\mbox{.}(2023)]%
        {albaladejo2023evaluating}
\bibfield{author}{\bibinfo{person}{Mariano Albaladejo-Gonz{\'a}lez}, \bibinfo{person}{Jos{\'e}~A Ruip{\'e}rez-Valiente}, {and} \bibinfo{person}{F{\'e}lix G{\'o}mez~M{\'a}rmol}.} \bibinfo{year}{2023}\natexlab{}.
\newblock \showarticletitle{Evaluating different configurations of machine learning models and their transfer learning capabilities for stress detection using heart rate}.
\newblock \bibinfo{journal}{\emph{Journal of Ambient Intelligence and Humanized Computing}} \bibinfo{volume}{14}, \bibinfo{number}{8} (\bibinfo{year}{2023}), \bibinfo{pages}{11011--11021}.
\newblock


\bibitem[Alberdi et~al\mbox{.}(2016)]%
        {alberdi2016towards}
\bibfield{author}{\bibinfo{person}{Ane Alberdi}, \bibinfo{person}{Asier Aztiria}, {and} \bibinfo{person}{Adrian Basarab}.} \bibinfo{year}{2016}\natexlab{}.
\newblock \showarticletitle{Towards an automatic early stress recognition system for office environments based on multimodal measurements: A review}.
\newblock \bibinfo{journal}{\emph{Journal of biomedical informatics}}  \bibinfo{volume}{59} (\bibinfo{year}{2016}), \bibinfo{pages}{49--75}.
\newblock


\bibitem[Baird et~al\mbox{.}(2021)]%
        {baird2021evaluation}
\bibfield{author}{\bibinfo{person}{Alice Baird}, \bibinfo{person}{Andreas Triantafyllopoulos}, \bibinfo{person}{Sandra Z{\"a}nkert}, \bibinfo{person}{Sandra Ottl}, \bibinfo{person}{Lukas Christ}, \bibinfo{person}{Lukas Stappen}, \bibinfo{person}{Julian Konzok}, \bibinfo{person}{Sarah Sturmbauer}, \bibinfo{person}{Eva-Maria Me{\ss}ner}, \bibinfo{person}{Brigitte~M Kudielka}, {et~al\mbox{.}}} \bibinfo{year}{2021}\natexlab{}.
\newblock \showarticletitle{An evaluation of speech-based recognition of emotional and physiological markers of stress}.
\newblock \bibinfo{journal}{\emph{Frontiers in Computer Science}}  \bibinfo{volume}{3} (\bibinfo{year}{2021}), \bibinfo{pages}{750284}.
\newblock


\bibitem[Balcombe and De~Leo(2022)]%
        {balcombe2022human}
\bibfield{author}{\bibinfo{person}{Luke Balcombe} {and} \bibinfo{person}{Diego De~Leo}.} \bibinfo{year}{2022}\natexlab{}.
\newblock \showarticletitle{Human-computer interaction in digital mental health}. In \bibinfo{booktitle}{\emph{Informatics}}, Vol.~\bibinfo{volume}{9}. MDPI, \bibinfo{pages}{14}.
\newblock


\bibitem[Benchekroun et~al\mbox{.}(2022)]%
        {benchekroun2022multi}
\bibfield{author}{\bibinfo{person}{Mouna Benchekroun}, \bibinfo{person}{Dan Istrate}, \bibinfo{person}{Vincent Zalc}, {and} \bibinfo{person}{Dominique Lenne}.} \bibinfo{year}{2022}\natexlab{}.
\newblock \showarticletitle{A Multi-Modal Dataset (MMSD) for Acute Stress Bio-Markers}. In \bibinfo{booktitle}{\emph{International Joint Conference on Biomedical Engineering Systems and Technologies}}. Springer, \bibinfo{pages}{377--392}.
\newblock


\bibitem[Benchekroun et~al\mbox{.}(2023)]%
        {benchekroun2023cross}
\bibfield{author}{\bibinfo{person}{Mouna Benchekroun}, \bibinfo{person}{Pedro~Elkind Velmovitsky}, \bibinfo{person}{Dan Istrate}, \bibinfo{person}{Vincent Zalc}, \bibinfo{person}{Plinio~Pelegrini Morita}, {and} \bibinfo{person}{Dominique Lenne}.} \bibinfo{year}{2023}\natexlab{}.
\newblock \showarticletitle{Cross dataset analysis for generalizability of HRV-based stress detection models}.
\newblock \bibinfo{journal}{\emph{Sensors}} \bibinfo{volume}{23}, \bibinfo{number}{4} (\bibinfo{year}{2023}), \bibinfo{pages}{1807}.
\newblock


\bibitem[Birjandtalab et~al\mbox{.}(2016)]%
        {birjandtalab2016non}
\bibfield{author}{\bibinfo{person}{Javad Birjandtalab}, \bibinfo{person}{Diana Cogan}, \bibinfo{person}{Maziyar~Baran Pouyan}, {and} \bibinfo{person}{Mehrdad Nourani}.} \bibinfo{year}{2016}\natexlab{}.
\newblock \showarticletitle{A non-EEG biosignals dataset for assessment and visualization of neurological status}. In \bibinfo{booktitle}{\emph{2016 IEEE International Workshop on Signal Processing Systems (SiPS)}}. IEEE, \bibinfo{pages}{110--114}.
\newblock


\bibitem[Bobade and Vani(2020)]%
        {bobade2020stress}
\bibfield{author}{\bibinfo{person}{Pramod Bobade} {and} \bibinfo{person}{M Vani}.} \bibinfo{year}{2020}\natexlab{}.
\newblock \showarticletitle{Stress detection with machine learning and deep learning using multimodal physiological data}. In \bibinfo{booktitle}{\emph{2020 Second International Conference on Inventive Research in Computing Applications (ICIRCA)}}. IEEE, \bibinfo{pages}{51--57}.
\newblock


\bibitem[Braithwaite et~al\mbox{.}(2013)]%
        {braithwaite2013guide}
\bibfield{author}{\bibinfo{person}{Jason~J Braithwaite}, \bibinfo{person}{Derrick~G Watson}, \bibinfo{person}{Robert Jones}, {and} \bibinfo{person}{Mickey Rowe}.} \bibinfo{year}{2013}\natexlab{}.
\newblock \showarticletitle{A guide for analysing electrodermal activity (EDA) \& skin conductance responses (SCRs) for psychological experiments}.
\newblock \bibinfo{journal}{\emph{Psychophysiology}} \bibinfo{volume}{49}, \bibinfo{number}{1} (\bibinfo{year}{2013}), \bibinfo{pages}{1017--1034}.
\newblock


\bibitem[Can et~al\mbox{.}(2019)]%
        {can2019stress}
\bibfield{author}{\bibinfo{person}{Yekta~Said Can}, \bibinfo{person}{Bert Arnrich}, {and} \bibinfo{person}{Cem Ersoy}.} \bibinfo{year}{2019}\natexlab{}.
\newblock \showarticletitle{Stress detection in daily life scenarios using smart phones and wearable sensors: A survey}.
\newblock \bibinfo{journal}{\emph{Journal of biomedical informatics}}  \bibinfo{volume}{92} (\bibinfo{year}{2019}), \bibinfo{pages}{103139}.
\newblock


\bibitem[Elgendi et~al\mbox{.}(2010)]%
        {elgendi2010frequency}
\bibfield{author}{\bibinfo{person}{Mohamed Elgendi}, \bibinfo{person}{Mirjam Jonkman}, {and} \bibinfo{person}{Friso De~Boer}.} \bibinfo{year}{2010}\natexlab{}.
\newblock \showarticletitle{Frequency Bands Effects on QRS Detection.}
\newblock \bibinfo{journal}{\emph{Biosignals}}  \bibinfo{volume}{2003} (\bibinfo{year}{2010}), \bibinfo{pages}{2002}.
\newblock


\bibitem[Elgendi et~al\mbox{.}(2013)]%
        {elgendi2013systolic}
\bibfield{author}{\bibinfo{person}{Mohamed Elgendi}, \bibinfo{person}{Ian Norton}, \bibinfo{person}{Matt Brearley}, \bibinfo{person}{Derek Abbott}, {and} \bibinfo{person}{Dale Schuurmans}.} \bibinfo{year}{2013}\natexlab{}.
\newblock \showarticletitle{Systolic peak detection in acceleration photoplethysmograms measured from emergency responders in tropical conditions}.
\newblock \bibinfo{journal}{\emph{PloS one}} \bibinfo{volume}{8}, \bibinfo{number}{10} (\bibinfo{year}{2013}), \bibinfo{pages}{e76585}.
\newblock


\bibitem[Giannakakis et~al\mbox{.}(2019)]%
        {giannakakis2019review}
\bibfield{author}{\bibinfo{person}{Giorgos Giannakakis}, \bibinfo{person}{Dimitris Grigoriadis}, \bibinfo{person}{Katerina Giannakaki}, \bibinfo{person}{Olympia Simantiraki}, \bibinfo{person}{Alexandros Roniotis}, {and} \bibinfo{person}{Manolis Tsiknakis}.} \bibinfo{year}{2019}\natexlab{}.
\newblock \showarticletitle{Review on psychological stress detection using biosignals}.
\newblock \bibinfo{journal}{\emph{IEEE transactions on affective computing}} \bibinfo{volume}{13}, \bibinfo{number}{1} (\bibinfo{year}{2019}), \bibinfo{pages}{440--460}.
\newblock


\bibitem[Greene et~al\mbox{.}(2016)]%
        {greene2016survey}
\bibfield{author}{\bibinfo{person}{Shalom Greene}, \bibinfo{person}{Himanshu Thapliyal}, {and} \bibinfo{person}{Allison Caban-Holt}.} \bibinfo{year}{2016}\natexlab{}.
\newblock \showarticletitle{A survey of affective computing for stress detection: Evaluating technologies in stress detection for better health}.
\newblock \bibinfo{journal}{\emph{IEEE Consumer Electronics Magazine}} \bibinfo{volume}{5}, \bibinfo{number}{4} (\bibinfo{year}{2016}), \bibinfo{pages}{44--56}.
\newblock


\bibitem[Gupta et~al\mbox{.}(2023)]%
        {gupta2023multimodal}
\bibfield{author}{\bibinfo{person}{Rohit Gupta}, \bibinfo{person}{Amit Bhongade}, {and} \bibinfo{person}{Tapan~Kumar Gandhi}.} \bibinfo{year}{2023}\natexlab{}.
\newblock \showarticletitle{Multimodal Wearable Sensors-based Stress and Affective States Prediction Model}. In \bibinfo{booktitle}{\emph{2023 9th International Conference on Advanced Computing and Communication Systems (ICACCS)}}, Vol.~\bibinfo{volume}{1}. IEEE, \bibinfo{pages}{30--35}.
\newblock


\bibitem[Haque et~al\mbox{.}(2024)]%
        {haque2024state}
\bibfield{author}{\bibinfo{person}{Yeaminul Haque}, \bibinfo{person}{Rahat~Shahriar Zawad}, \bibinfo{person}{Chowdhury Saleh~Ahmed Rony}, \bibinfo{person}{Hasan Al~Banna}, \bibinfo{person}{Tapotosh Ghosh}, \bibinfo{person}{M~Shamim Kaiser}, {and} \bibinfo{person}{Mufti Mahmud}.} \bibinfo{year}{2024}\natexlab{}.
\newblock \showarticletitle{State-of-the-Art of Stress Prediction from Heart Rate Variability Using Artificial Intelligence}.
\newblock \bibinfo{journal}{\emph{Cognitive Computation}} \bibinfo{volume}{16}, \bibinfo{number}{2} (\bibinfo{year}{2024}), \bibinfo{pages}{455--481}.
\newblock


\bibitem[Heimerl et~al\mbox{.}(2023)]%
        {heimerl2023fordigitstress}
\bibfield{author}{\bibinfo{person}{Alexander Heimerl}, \bibinfo{person}{Pooja Prajod}, \bibinfo{person}{Silvan Mertes}, \bibinfo{person}{Tobias Baur}, \bibinfo{person}{Matthias Kraus}, \bibinfo{person}{Ailin Liu}, \bibinfo{person}{Helen Risack}, \bibinfo{person}{Nicolas Rohleder}, \bibinfo{person}{Elisabeth Andr{\'e}}, {and} \bibinfo{person}{Linda Becker}.} \bibinfo{year}{2023}\natexlab{}.
\newblock \showarticletitle{ForDigitStress: A multi-modal stress dataset employing a digital job interview scenario}.
\newblock \bibinfo{journal}{\emph{arXiv preprint arXiv:2303.07742}} (\bibinfo{year}{2023}).
\newblock


\bibitem[Koldijk et~al\mbox{.}(2016)]%
        {koldijk2016detecting}
\bibfield{author}{\bibinfo{person}{Saskia Koldijk}, \bibinfo{person}{Mark~A Neerincx}, {and} \bibinfo{person}{Wessel Kraaij}.} \bibinfo{year}{2016}\natexlab{}.
\newblock \showarticletitle{Detecting work stress in offices by combining unobtrusive sensors}.
\newblock \bibinfo{journal}{\emph{IEEE Transactions on affective computing}} \bibinfo{volume}{9}, \bibinfo{number}{2} (\bibinfo{year}{2016}), \bibinfo{pages}{227--239}.
\newblock


\bibitem[Koldijk et~al\mbox{.}(2014)]%
        {koldijk2014swell}
\bibfield{author}{\bibinfo{person}{Saskia Koldijk}, \bibinfo{person}{Maya Sappelli}, \bibinfo{person}{Suzan Verberne}, \bibinfo{person}{Mark~A Neerincx}, {and} \bibinfo{person}{Wessel Kraaij}.} \bibinfo{year}{2014}\natexlab{}.
\newblock \showarticletitle{The SWELL knowledge work dataset for stress and user modeling research}. In \bibinfo{booktitle}{\emph{Proceedings of the 16th international conference on multimodal interaction}}. \bibinfo{pages}{291--298}.
\newblock


\bibitem[Liapis et~al\mbox{.}(2021)]%
        {liapis2021detection}
\bibfield{author}{\bibinfo{person}{Alexandros Liapis}, \bibinfo{person}{Evanthia Faliagka}, \bibinfo{person}{Christos Katsanos}, \bibinfo{person}{Christos Antonopoulos}, {and} \bibinfo{person}{Nikolaos Voros}.} \bibinfo{year}{2021}\natexlab{}.
\newblock \showarticletitle{Detection of subtle stress episodes during UX evaluation: Assessing the performance of the WESAD bio-signals dataset}. In \bibinfo{booktitle}{\emph{Human-Computer Interaction--INTERACT 2021: 18th IFIP TC 13 International Conference, Bari, Italy, August 30--September 3, 2021, Proceedings, Part III 18}}. Springer, \bibinfo{pages}{238--247}.
\newblock


\bibitem[Luong et~al\mbox{.}(2020)]%
        {luong2020towards}
\bibfield{author}{\bibinfo{person}{Tiffany Luong}, \bibinfo{person}{Nicolas Martin}, \bibinfo{person}{Anais Raison}, \bibinfo{person}{Ferran Argelaguet}, \bibinfo{person}{Jean-Marc Diverrez}, {and} \bibinfo{person}{Anatole L{\'e}cuyer}.} \bibinfo{year}{2020}\natexlab{}.
\newblock \showarticletitle{Towards real-time recognition of users mental workload using integrated physiological sensors into a VR HMD}. In \bibinfo{booktitle}{\emph{2020 IEEE international symposium on mixed and augmented reality (ISMAR)}}. IEEE, \bibinfo{pages}{425--437}.
\newblock


\bibitem[Makowski et~al\mbox{.}(2021)]%
        {makowski2021neurokit2}
\bibfield{author}{\bibinfo{person}{Dominique Makowski}, \bibinfo{person}{Tam Pham}, \bibinfo{person}{Zen~J Lau}, \bibinfo{person}{Jan~C Brammer}, \bibinfo{person}{Fran{\c{c}}ois Lespinasse}, \bibinfo{person}{Hung Pham}, \bibinfo{person}{Christopher Sch{\"o}lzel}, {and} \bibinfo{person}{SH~Annabel Chen}.} \bibinfo{year}{2021}\natexlab{}.
\newblock \showarticletitle{NeuroKit2: A Python toolbox for neurophysiological signal processing}.
\newblock \bibinfo{journal}{\emph{Behavior research methods}} (\bibinfo{year}{2021}), \bibinfo{pages}{1--8}.
\newblock


\bibitem[Martinho et~al\mbox{.}(2018)]%
        {martinho2018towards}
\bibfield{author}{\bibinfo{person}{Miguel Martinho}, \bibinfo{person}{Ana Fred}, {and} \bibinfo{person}{Hugo Silva}.} \bibinfo{year}{2018}\natexlab{}.
\newblock \showarticletitle{Towards continuous user recognition by exploring physiological multimodality: An electrocardiogram (ECG) and blood volume pulse (BVP) approach}. In \bibinfo{booktitle}{\emph{2018 International Symposium in Sensing and Instrumentation in IoT Era (ISSI)}}. IEEE, \bibinfo{pages}{1--6}.
\newblock


\bibitem[Mishra et~al\mbox{.}(2020)]%
        {mishra2020evaluating}
\bibfield{author}{\bibinfo{person}{Varun Mishra}, \bibinfo{person}{Sougata Sen}, \bibinfo{person}{Grace Chen}, \bibinfo{person}{Tian Hao}, \bibinfo{person}{Jeffrey Rogers}, \bibinfo{person}{Ching-Hua Chen}, {and} \bibinfo{person}{David Kotz}.} \bibinfo{year}{2020}\natexlab{}.
\newblock \showarticletitle{Evaluating the reproducibility of physiological stress detection models}.
\newblock \bibinfo{journal}{\emph{Proceedings of the ACM on interactive, mobile, wearable and ubiquitous technologies}} \bibinfo{volume}{4}, \bibinfo{number}{4} (\bibinfo{year}{2020}), \bibinfo{pages}{1--29}.
\newblock


\bibitem[Nkurikiyeyezu et~al\mbox{.}(2019)]%
        {nkurikiyeyezu2019effect}
\bibfield{author}{\bibinfo{person}{Kizito Nkurikiyeyezu}, \bibinfo{person}{Anna Yokokubo}, {and} \bibinfo{person}{Guillaume Lopez}.} \bibinfo{year}{2019}\natexlab{}.
\newblock \showarticletitle{The effect of person-specific biometrics in improving generic stress predictive models}.
\newblock \bibinfo{journal}{\emph{arXiv preprint arXiv:1910.01770}} (\bibinfo{year}{2019}).
\newblock


\bibitem[Pham et~al\mbox{.}(2021)]%
        {pham2021heart}
\bibfield{author}{\bibinfo{person}{Tam Pham}, \bibinfo{person}{Zen~Juen Lau}, \bibinfo{person}{SH~Annabel Chen}, {and} \bibinfo{person}{Dominique Makowski}.} \bibinfo{year}{2021}\natexlab{}.
\newblock \showarticletitle{Heart rate variability in psychology: A review of HRV indices and an analysis tutorial}.
\newblock \bibinfo{journal}{\emph{Sensors}} \bibinfo{volume}{21}, \bibinfo{number}{12} (\bibinfo{year}{2021}), \bibinfo{pages}{3998}.
\newblock


\bibitem[Prajod and Andr{\'e}(2022)]%
        {prajod2022generalizability}
\bibfield{author}{\bibinfo{person}{Pooja Prajod} {and} \bibinfo{person}{Elisabeth Andr{\'e}}.} \bibinfo{year}{2022}\natexlab{}.
\newblock \showarticletitle{On the Generalizability of ECG-based Stress Detection Models}. In \bibinfo{booktitle}{\emph{2022 21st IEEE International Conference on Machine Learning and Applications (ICMLA)}}. IEEE, \bibinfo{pages}{549--554}.
\newblock


\bibitem[Prajod et~al\mbox{.}({[n.\,d.]})]%
        {prajod11flow}
\bibfield{author}{\bibinfo{person}{Pooja Prajod}, \bibinfo{person}{Matteo Lavit~Nicora}, \bibinfo{person}{Marta Mondellini}, \bibinfo{person}{Matteo~Meregalli Falerni}, \bibinfo{person}{Rocco Vertechy}, \bibinfo{person}{Matteo Malosio}, {and} \bibinfo{person}{Elisabeth Andr{\'e}}.} \bibinfo{year}{[n.\,d.]}\natexlab{}.
\newblock \showarticletitle{Flow in Human-Robot Collaboration-Multimodal Analysis and Perceived Challenge Detection in Industrial Scenarios}.
\newblock \bibinfo{journal}{\emph{Frontiers in Robotics and AI}}  \bibinfo{volume}{11} (\bibinfo{year}{[n.\,d.]}), \bibinfo{pages}{1393795}.
\newblock


\bibitem[Sabour et~al\mbox{.}(2021)]%
        {sabour2021ubfc}
\bibfield{author}{\bibinfo{person}{Rita~Meziati Sabour}, \bibinfo{person}{Yannick Benezeth}, \bibinfo{person}{Pierre De~Oliveira}, \bibinfo{person}{Julien Chappe}, {and} \bibinfo{person}{Fan Yang}.} \bibinfo{year}{2021}\natexlab{}.
\newblock \showarticletitle{Ubfc-phys: A multimodal database for psychophysiological studies of social stress}.
\newblock \bibinfo{journal}{\emph{IEEE Transactions on Affective Computing}} \bibinfo{volume}{14}, \bibinfo{number}{1} (\bibinfo{year}{2021}), \bibinfo{pages}{622--636}.
\newblock


\bibitem[Sarkar and Etemad(2020)]%
        {sarkar2020self}
\bibfield{author}{\bibinfo{person}{Pritam Sarkar} {and} \bibinfo{person}{Ali Etemad}.} \bibinfo{year}{2020}\natexlab{}.
\newblock \showarticletitle{Self-supervised ECG representation learning for emotion recognition}.
\newblock \bibinfo{journal}{\emph{IEEE Transactions on Affective Computing}} \bibinfo{volume}{13}, \bibinfo{number}{3} (\bibinfo{year}{2020}), \bibinfo{pages}{1541--1554}.
\newblock


\bibitem[Schmidt et~al\mbox{.}(2018)]%
        {schmidt2018introducing}
\bibfield{author}{\bibinfo{person}{Philip Schmidt}, \bibinfo{person}{Attila Reiss}, \bibinfo{person}{Robert Duerichen}, \bibinfo{person}{Claus Marberger}, {and} \bibinfo{person}{Kristof Van~Laerhoven}.} \bibinfo{year}{2018}\natexlab{}.
\newblock \showarticletitle{Introducing WESAD, a multimodal dataset for wearable stress and affect detection}. In \bibinfo{booktitle}{\emph{Proceedings of the 20th ACM international conference on multimodal interaction}}. \bibinfo{pages}{400--408}.
\newblock


\bibitem[Velmovitsky et~al\mbox{.}(2021)]%
        {velmovitsky2021towards}
\bibfield{author}{\bibinfo{person}{Pedro~Elkind Velmovitsky}, \bibinfo{person}{Paulo Alencar}, \bibinfo{person}{Scott~T Leatherdale}, \bibinfo{person}{Donald Cowan}, {and} \bibinfo{person}{Plinio~Pelegrini Morita}.} \bibinfo{year}{2021}\natexlab{}.
\newblock \showarticletitle{Towards real-time public health: a novel mobile health monitoring system}. In \bibinfo{booktitle}{\emph{2021 IEEE International Conference on Big Data (Big Data)}}. IEEE, \bibinfo{pages}{6049--6051}.
\newblock


\bibitem[Vos et~al\mbox{.}(2023a)]%
        {vos2023ensemble}
\bibfield{author}{\bibinfo{person}{Gideon Vos}, \bibinfo{person}{Kelly Trinh}, \bibinfo{person}{Zoltan Sarnyai}, {and} \bibinfo{person}{Mostafa~Rahimi Azghadi}.} \bibinfo{year}{2023}\natexlab{a}.
\newblock \showarticletitle{Ensemble machine learning model trained on a new synthesized dataset generalizes well for stress prediction using wearable devices}.
\newblock \bibinfo{journal}{\emph{Journal of Biomedical Informatics}}  \bibinfo{volume}{148} (\bibinfo{year}{2023}), \bibinfo{pages}{104556}.
\newblock


\bibitem[Vos et~al\mbox{.}(2023b)]%
        {vos2023generalizable}
\bibfield{author}{\bibinfo{person}{Gideon Vos}, \bibinfo{person}{Kelly Trinh}, \bibinfo{person}{Zoltan Sarnyai}, {and} \bibinfo{person}{Mostafa~Rahimi Azghadi}.} \bibinfo{year}{2023}\natexlab{b}.
\newblock \showarticletitle{Generalizable machine learning for stress monitoring from wearable devices: a systematic literature review}.
\newblock \bibinfo{journal}{\emph{International Journal of Medical Informatics}}  \bibinfo{volume}{173} (\bibinfo{year}{2023}), \bibinfo{pages}{105026}.
\newblock


\bibitem[Yadav et~al\mbox{.}(2020)]%
        {yadav2020exploring}
\bibfield{author}{\bibinfo{person}{Megha Yadav}, \bibinfo{person}{Md~Nazmus Sakib}, \bibinfo{person}{Ehsanul~Haque Nirjhar}, \bibinfo{person}{Kexin Feng}, \bibinfo{person}{Amir~H Behzadan}, {and} \bibinfo{person}{Theodora Chaspari}.} \bibinfo{year}{2020}\natexlab{}.
\newblock \showarticletitle{Exploring individual differences of public speaking anxiety in real-life and virtual presentations}.
\newblock \bibinfo{journal}{\emph{IEEE Transactions on Affective Computing}} \bibinfo{volume}{13}, \bibinfo{number}{3} (\bibinfo{year}{2020}), \bibinfo{pages}{1168--1182}.
\newblock


\bibitem[Yu et~al\mbox{.}(2018)]%
        {yu2018biofeedback}
\bibfield{author}{\bibinfo{person}{Bin Yu}, \bibinfo{person}{Mathias Funk}, \bibinfo{person}{Jun Hu}, \bibinfo{person}{Qi Wang}, {and} \bibinfo{person}{Loe Feijs}.} \bibinfo{year}{2018}\natexlab{}.
\newblock \showarticletitle{Biofeedback for everyday stress management: A systematic review}.
\newblock \bibinfo{journal}{\emph{Frontiers in ICT}}  \bibinfo{volume}{5} (\bibinfo{year}{2018}), \bibinfo{pages}{23}.
\newblock


\bibitem[Zawad et~al\mbox{.}(2023)]%
        {zawad2023hybrid}
\bibfield{author}{\bibinfo{person}{Md~Rahat~Shahriar Zawad}, \bibinfo{person}{Chowdhury Saleh~Ahmed Rony}, \bibinfo{person}{Md~Yeaminul Haque}, \bibinfo{person}{Md~Hasan~Al Banna}, \bibinfo{person}{Mufti Mahmud}, {and} \bibinfo{person}{M~Shamim Kaiser}.} \bibinfo{year}{2023}\natexlab{}.
\newblock \showarticletitle{A Hybrid Approach for Stress Prediction from Heart Rate Variability}.
\newblock In \bibinfo{booktitle}{\emph{Frontiers of ICT in Healthcare: Proceedings of EAIT 2022}}. \bibinfo{publisher}{Springer}, \bibinfo{pages}{111--121}.
\newblock


\end{thebibliography}

\end{document}